\documentclass[useAMS, referee]{biom}
\def\bSig\mathbf{\Sigma}

\usepackage{xcolor}
\usepackage{amsmath,amssymb}
\usepackage{enumerate, gensymb}
\usepackage{float}
\usepackage{lipsum}
\usepackage{mathrsfs}
\usepackage{adjustbox}
\usepackage{array,multirow,graphicx}
\usepackage{subfig}

\let\bibhang\relax
\usepackage{natbib}
\usepackage{booktabs}

\title[The Robust Kernel Association Test (RobKAT)]{The Robust Kernel Association Test (RobKAT) }

\author{Kara Martinez$^{1}$, 
Arnab Maity$^{1,*}$\email{amaity@ncsu.edu}, Robert H. Yolken$^{5}$, Patrick F. Sullivan$^{6}$, and Jung-Ying Tzeng$^{1,2,3,4}$ \\
$^{1}$Department of Statistics, North Carolina State University, Raleigh, North Carolina, U.S.A. \\
$^{2}$Bioinformatics Research Center, North Carolina State University, Raleigh, North Carolina, U.S.A. \\
$^{3}$Institute of Epidemiology and Preventive Medicine, National Taiwan University, Taipei, Taiwan \\
$^{4}$Department of Statistics, National Cheng-Kung University, Tainan City, Taiwan \\
$^{5}$Stanley Neurovirology Laboratory, Johns Hopkins School of Medicine, USA.\\
$^{6}$Department of Genetics, University of North Carolina at Chapel Hill, Chapel Hill, North Carolina, U.S.A }

\begin{document}


\date{{\it Received MONTH} YEAR. {\it Revised MONTH} YEAR.  {\it
Accepted MONTH} YEAR.}



\pagerange{\pageref{firstpage}--\pageref{lastpage}} 
\volume{00}
\pubyear{0000}
\artmonth{000}


\doi{000}


\label{firstpage}


\begin{abstract}
Testing the association between SNP effects and a response is a common task. Such tests are often carried out through kernel machine methods based on least squares, such as the Sequence Kernel Association Test (SKAT). However, these least squares procedures assume a normally distributed response, which is often violated. Other robust procedures such as the Quantile Regression Kernel Machine (QRKM) restrict choice of loss function and only allow inference on conditional quantiles. We propose a general and robust kernel association test with flexible choice of loss function, no distributional assumptions, and has SKAT and QRKM as special cases. We evaluate our proposed robust association test (RobKAT) across various data distributions through simulation study. When errors are normally distributed, RobKAT controls type I error and shows comparable power to SKAT. In all other distributional settings investigated, our robust test has similar or greater power than SKAT. Finally, we apply our robust kernel association test on data from the CATIE clinical trial to detect associations between selected genes on chromosome 6, including the Major Histocompatibility Complex (MHC) region, and neurotrophic herpesvirus antibody levels in schizophrenia patients. RobKAT detected significant association with four SNP-sets (HST1H2BJ, MHC, POM12L2, and SLC17A1), three of which were undetected by SKAT.
\end{abstract}

%



\maketitle


%

\section{Introduction}
\label{s:intro}

Detecting associations between a set of genetic markers, such as SNPs, and a continuous response is a common problem. These tasks lend themselves to multi-marker approaches, such as kernel methods, that analyze variants at the gene level and may gain power due to their lower multiple testing burden. Kernel methods are also appealing over burden-based multi-marker approaches, which treat the overall genetic marker effect as fixed and test for the mean genetic effect with various weighting schemes \citep{Gauderman2007, Madsen2009, Li2008}. Many of these burden-based methods are optimal in the case of additive loci with similar size and direction of effect, but kernel methods typically avoid these loci effect restrictions by viewing the overall genetic effect as random or the output of a nonparametric function.
			
Least squares kernel machines (LSKM), developed by \citet{Liu2007} and \citet{Kwee2008}, are popular kernel methods that utilize a linear mixed model with normally distributed errors. Mixed model methods such as Restricted Maximum Likelihood (REML) may be applied for simultaneous estimation of parameters, and LSKM estimators can be shown to be the best linear unbiased predictors. Score tests can be used for testing the overall genetic marker effect, and due to distributional assumptions this score statistic follows a Chi-squared mixture under the null hypothesis \citep{Liu2007}. A type of LSKM that has been developed is the sequence kernel association test (SKAT), which incorporates weights into the kernel according to genetic marker minor allele frequencies \citep{Wu2011}. However, the assumption of normally distributed errors limits inference to the conditional mean response. In the case of a skewed or heavy-tailed response, the null distribution of the test statistic does not follow a Chi-squared mixture, leading to moderate to severe loss of power.
			
A variation on LSKMs, the quantile regression kernel machines (QRKM) utilizes quantile regression methods rather than least squares for parameter estimation. When testing the genetic marker effect, \citet{Kong2016} proposes a QRKM score-type statistic based on the subgradient of the check function rather than the traditional score test. Although QRKM is more robust to extreme observations, the method is restricted only to inference on the conditional quantile. While these quantiles offer more choice than in LSKMs, additional flexibility may be desired.  

Although appropriate multi-marker association testing procedures are limited for non-normal responses, one such data set is the the Clinical Antipsychotic Trials of Intervention Effectiveness (CATIE) study. The CATIE trial aimed to compare the effectiveness of atypical antipsychotics in unrelated patients suffering from schizophrenia and measured patient antibody levels on three neurotrophic herpesviruses. As discussed with additional detail in Section \ref{s:dataAnalysis}, these antibody levels are notably right-skewed. Even after a log-transformation, the data display a bimodal distribution. Violating the assumption of normality, tests such as LSKM and SKAT are not appropriate. Although QRKM is valid with such distributions and robust to outliers, QRKM restricts inference to the conditional quantile and limits the loss function used in the analysis. Thus, we propose a robust and general method to test for multi-marker association with this non-normally distributed response.   
      
Our robust kernel association test accommodates any valid loss function and is robust to outliers, heavy tails, and skewed distributions. We achieve this robust framework by generalizing previous kernel methods and eliminating all distributional assumptions on the response. While LSKM and SKAT specify a square loss and QRKM specifies a 'check' loss, we follow similar score function derivations with a general loss function subject to a few minor restrictions. The user-specified, general loss function may be taken from the robust statistics literature, rendering RobKAT methodology robust to non-normal responses. The RobKAT test statistic also is a similar quadratic form as previous methods and, as described in Section \ref{s:methodology}, simplifies to LSKM, SKAT, and QRKM in special cases. However, RobKAT's general loss function also provides alternatives to LSKM's conditional mean inference and QRKM's conditional quantile inference. The null distribution of our test statistic and associated p-values can be calculated through a permutation test, as originally proposed for QRKM. However, since the permutation testing procedure is computationally inefficient, we instead apply a fast permutation testing method described by \citet{Zhan2017} that estimates the null distribution of the test statistic through moment matching.

The rest of the article is organized as follows. Section \ref{s:methodology} details model notation, derivation of our RobKAT statistic, and p-value computation. Section \ref{s:simulation} then demonstrates the validity, Type I error control, and power of our test through simulation study comparing to normal-based methods. We show that our proposed testing procedure is robust to non-normal distributions, even providing a powerful and Type I error-controlled test when errors follow a Cauchy distribution. Although comparable to SKAT when the response is normal, RobKAT shows greater power in all non-normal cases that we explored. In Section \ref{s:dataAnalysis}, we apply our proposed testing procedure to detect the association between genes in chromosome 6, including the MHC region, and antibody levels using CATIE clinical trial data. We find four significantly associated gene sets using the skewed antibody response, three of which are not detected by SKAT. In Section \ref{s:discuss}, we conclude that RobKAT's computational efficiency and robust nature make it a useful procedure even on the genome-wide scale.



\section{Methodology}
\label{s:methodology}

\subsection{Model Specification and Testing Problem}

Suppose we observe data $(Y_i, \bmath{X}_i, \bmath{Z_i}), i = 1, \ldots, n$, where $Y_i$ is a scalar response of interest, $\bmath{Z}_i$ denotes a $p\times 1$ vector of genetic covariates, and $\bmath{X}_i$ denotes a $q \times 1$ vector of other confounders observed for the $i$th subject.  We consider the model \[Y_i = \bmath{X}_i^T \bmath{\beta} + h(\bmath{Z}_i) + \epsilon_i,\]
where $\bmath{\beta}$ is an unknown coefficient vector, $h(\cdot)$ is an unknown function quantifying the effect of the genetic covariates, and $\epsilon_i, i = 1, \ldots, n$ are independent mean zero random errors. Testing for the overall genetic marker effect corresponds to a hypothesis test for $H_0: h(\bmath{Z}_i) = 0$.

One issue in kernel machine regression is that the potentially nonparametric function $h(\cdot)$ may be difficult to model. We instead assume that $h(\cdot)$ is an element of a reproducing kernel Hilbert space (RKHS), $\mathcal{H}_K$, which can be uniquely defined by a positive semi-definite kernel function $k(\cdot, \cdot)$ \citep{Shawe2004, Wahba1990}. Due to certain mathematical properties of RKHS, this structure allows us to represent $h(\bmath{Z}_i)$ with images of the kernel function. Further, we may summarize this information in the positive semi-definite matrix $\mathbb{K}$, whose $(i,j)$ element is defined by $k(\bmath{Z}_i, \bmath{Z}_j)$. To obtain estimates of $h(\cdot)$, researchers often minimize the penalized loss criterion \begin{equation} \label{minCriterion} \min_{\bmath{\beta}, h\in \mathcal{H}_K} \ \bigg[ \sum_{i=1}^N\rho\{Y_i - \bmath{X}_i^T \bmath{\beta} - h(\bmath{Z}_i)\} + \lambda \ ||h||^2_{\mathcal{H}_K} \bigg] \end{equation} We note that the nonparametric functional form of $h(\cdot)$ may be infinite-dimensional. Regarding the form of $\rho(\cdot)$, the least squares kernel machine methods specify $\rho(t) = t^2/2$ in order to implement squared-loss. 

A key observation is that estimates of $\widehat{\bmath{\beta}}$ and $\widehat{\bmath{h}}$ from Equation \ref{minCriterion} match best linear unbiased estimators estimates obtained from the mixed effects model, $Y_i = \bmath{X}_i^T \bmath{\beta} + h_i + \epsilon_i$ where $\bmath{h} \sim N(0, \tau \mathbb{K})$ and $\epsilon_i \sim N(0, \sigma^2)$ \citep{Liu2007}. Further, estimates for the remaining parameters, $\tau$ and $\sigma^2$, can be computed using restricted maximum likelihood estimates. With the mixed effects model in mind, LSKM performs a hypothesis test for $H_0: \tau = 0$ using a variance component score test. Leveraging the model's normality assumptions, inference on the test statistic comes from a Chi-Squared mixture and a Satterthwaite approximation \citep{Liu2007}.

QRKM also uses the penalized minimization criterion but instead specifies a check function for the loss: $\rho(t) = \eta t \ \mathbb{I}_{\{t \geq 0\}} + (\eta - 1)t \ \mathbb{I}_{\{t < 0\}}$ where $\eta \in (0,1)$ is the conditional quantile of interest \citep{Kong2016}. When researchers are interested in association with the conditional median response, for example, $\eta = 0.5$ and $\rho(t) = 0.5 |t|$. Estimates for $\bmath{\beta}$ are the values that minimize $\sum_{i=1}^N\rho(Y_i - X_i^T \bmath{\beta})$, Equation 1 under the null hypothesis, and can be computed using quantile regression packages such as \verb|quantreg| in R. As with LSKM, the QRKM test statistic follows the variance component score test procedure. Since the check loss function is not differentiable, however, the derivation uses the subgradient, $\psi(t) = \eta \mathbb{I}_{\{t > 0\}} + (\eta - 1) \mathbb{I}_{\{t < 0\}} + (\eta - B)\mathbb{I}_{\{t = 0\}}$, where $B$ is the outcome of a $Bern(\eta)$ random variable. Without using an assumption of normality, the null distribution of the QRKM test statistic and resulting p-value are computed using a permutation algorithm \citep{Kong2016}.

In RobKAT, we derive the variance component score statistic but with a general $\rho$, subject only to a small number of conditions. In this way, LSKM and QRKM will be special cases of our robust and generalized method. Let $\rho$ be a differentiable function $\rho: \mathbb{R} \rightarrow \mathbb{R}^+$ such that 
\begin{enumerate}
	\item $\rho(0) = 0$
	\item $\rho(x)$ is a convex function of $x$
    \item $\psi(x) = \frac{d}{dx} \rho(x)$ exists and is non-decreasing
\end{enumerate}
The first assumption is taken from standard robust statistics procedures described in \citet{Maronna2006}, and the last two are modifications to exclude redescending $\psi$-functions. Violations of assumptions 2 and 3 can be accommodated, however performance of the resulting test may be suboptimal. The exclusion of these redescending $\psi$-functions and such performance impairment is discussed further in section \ref{s:discuss}. Note that we do not make any assumptions on the distribution of $e_i$. 

\subsection{Test Statistic}

To motivate our method, we first consider the simpler case in which $h(\bmath{Z}_i) = \bmath{Z}_i^T\bmath{\gamma}$. Testing $H_0: h(\bmath{Z}_i) = 0$ in this case is equivalent to testing $H_0: \bmath{\gamma} = 0$. A score-type test follows from minimizing the criterion \[L(\bmath{\beta}, \bmath{\gamma}) = \sum_{i=1}^n \rho\bigg(\frac{e_i}{s}\bigg) = \sum_{i=1}^n \rho\bigg( \frac{Y_i - \bmath{X}_i^T\bmath{\beta} - \bmath{Z}_i^T\bmath{\gamma}}{s}\bigg)  \] where $s$ is the scale parameter for the distribution of $e_i$. Differentiating and setting equal to zero, the corresponding score function then yields the M-estimation equations \[S(\bmath{\beta}, \bmath{\gamma}) = \sum_{i=1}^n\bmath{Z}_i \ \psi\bigg(\frac{e_i}{s}\bigg) = \sum_{i=1}^n \bmath{Z}_i \ \psi\bigg(\frac{Y_i - \bmath{X}_i^T\bmath{\beta} - \bmath{Z}_i^T\bmath{\gamma}}{s}\bigg) = 0 \]

Let $\widehat{\bmath{\beta}}$ be an estimate of $\bmath{\beta}$ under the null hypothesis $H_0: h(\bmath{Z}_i) = \bmath{Z}_i^T\bmath{\gamma} = 0$, and let $\widehat{e_i} = Y_i + \bmath{X}_i^T\widehat{\bmath{\beta}}$. Under $H_0$, fitting $\widehat{\bmath{\beta}}$ and accordingly $S(\widehat{\bmath{\beta}}, \bmath{\gamma}) |_{\bmath{\gamma} = 0}$ reduces to the simple task of fitting the linear regression model $Y_i = \bmath{X}_i^T\bmath{\beta} + e_i$, which does not require knowledge of the form of $h(\bmath{Z}_i)$ and can be done easily using standard robust regression methods. When obtaining an M-estimate of $s$, minimizing $\sum_{i=1}^n \rho(\frac{e_i}{s})$ can ensure that regression estimates are invariant. However, we desire an estimate of $s$ that is also robust. One solution is to use Huber's Proposal 2 robust estimate of scale, which solves \[\frac{1}{n-p}\sum_{i=1}^n \psi^2\big(\frac{Y_i - \bmath{X}_i^T\widehat{\bmath{\beta}}}{s}\big) = E_{\Phi}(\psi^2(\bmath{\epsilon}))\] where $\Phi$ denotes an expectation under the assumption of a standard normal error distribution \citep{PeterHuber1977,Schrader1980}. Huber's Proposal 2 estimate is preferred over other robust estimates of scale as it has desirable convergence properties under our assumption of convex $\rho$ \citep{Schrader1980}. 

We then derive the test statistic 
$T = S(\widehat{\bmath{\beta}}, \bmath{\gamma})^T S(\widehat{\bmath{\beta}}, \bmath{\gamma}) = \bmath{w}^T \mathbb{K} \bmath{w} $
The last expression uses the simplifications $\bmath{w}$ and $\mathbb{K}$ where the $i^{th}$ element of $\bmath{w}$ is defined by $w_i=\psi(\frac{\widehat{e_i}}{\widehat{s}})$ and the $(i,j)^{th}$ element of $\mathbb{K}$ is defined by $k(\bmath{Z}_i, \bmath{Z}_j) = \bmath{Z}_i^T \bmath{Z}_j$. 

This $\mathbb{K}$ can be generalized to any kernel function $k(.,.)$. The effect of $k(.,.)$ is linked with the form of $h(\bmath{Z}_i)$. In the previous derivation, setting $h(\bmath{Z}_i) = \bmath{Z}_i \bmath{\gamma}$, a model with only main genetic-marker effects, corresponds to the linear kernel $\mathbb{K}$ with entries defined by the function $k(\bmath{Z}_i,\bmath{Z}_j) = \bmath{Z}_i^T\bmath{Z}_j$. If instead for $\bmath{Z}_i \in \mathbb{R}^p$ we chose in our derivation $h(\bmath{Z}_i) = \sum_{j<k} \bmath{Z}_{ij}\bmath{Z}_{ik}\bmath{\gamma}_{jk} + \sum_{j=1}^p \bmath{Z}_{ij}^2\bmath{\gamma}_{2j}$, a model with quadratic main effects and all two-way interactions, then we would obtain a quadratic kernel whose entries are defined by $k(\bmath{Z}_i,\bmath{Z}_j) = (\bmath{Z}_i^T\bmath{Z}_j )^2$. The kernel function $k(.,.)$ generates a function space that contains $h(\bmath{Z}_i)$. This theoretical framework thus represents $h(\bmath{Z}_i)$ by a linear combination $h(\bmath{Z}_i) = \sum_j \alpha_j \ k(\bmath{Z}_i, \bmath{Z}_j)$ for some weights $\alpha_j$, which gives a clear connection between choice of $k(.,.)$ and form of $h(\bmath{Z}_i)$. Since $\mathbb{K}$ is symmetric positive semi-definite, $k(.,.)$ is often thought of as a similarity metric between two individuals, which may be more intuitive to specify than the form of $h(\bmath{Z}_i)$ itself. Further, any symmetric positive semi-definite matrix describing similarity may be used for $\mathbb{K}$. When $\bmath{Z}_i$ represents genetic markers, for example, the IBS kernel is defined as $k(\bmath{Z}_i, \bmath{Z}_j) = \frac{1}{2p}\sum_{k=1}^p [2\mathbb{I}\{\bmath{Z}_{ik} = \bmath{Z}_{jk}\} + \mathbb{I}\{|\bmath{Z}_{ik} - \bmath{Z}_{jk}| = 1\}]$ or the proportion of alleles shared between individuals $i$ and $j$.

By instead defining the form of $h(\bmath{Z}_i)$ through $k(.,.)$, we generalize our test statistic to any valid kernel function. Thus, we produce a test statistic that is intuitively specified through a kernel function, $k(.,.)$, and loss function, $\rho$. Note that any $\rho$ function following the general assumptions listed above is easily incorporated into this test statistic. For example, the case in which $\rho(t) = \frac{1}{2}t^2$ leads to the usual LSKM test statistic and $\rho(t) = 0.5|t|$ leads to the QRKM statistic \citep{Liu2007, Kong2016}. Another common choice of $\rho$ is from the family of Huber functions defined by \[\rho(x) = \begin{cases} \frac{1}{2}x^2 & |x|\leq k \\ k(|x|-\frac{k}{2}) & |x| > k \end{cases}\] where the value of $k$ may be chosen for a desired asymptotic variance \citep{PeterHuber1977,Maronna2006}. For $-k < x < k$, the function is quadratic as in least squares, and for $|x| > k$ the function is linear as in quantile regression. Thus, this family may be seen as an intermediate between least squares and quantile regression. 

To calculate our general kernel machine test statistic, we first fit the model under the null hypothesis, $H_0:  h(\bmath{Z}_i) = 0$, by regressing $\bmath{Y}$ on $\bmath{X}$ and  minimizing $\sum_{i=1}^n \rho(e_i)$ with the iterated re-weighted least squares algorithm. Then we use the null model to compute $\widehat{s}$, Huber’s Proposal 2 robust estimate of scale. We then compute the score-type statistic described above, $T = \bmath{w}^t \mathbb{K} \bmath{w}$. 

\subsection{Null Distribution of the Testing Statistic}
It is possible to obtain the null distribution for a statistic of the form $\bmath{w}^T \mathbb{K} \bmath{w}$ using permutations \citep{Kong2016}. However, there is a substantial computational burden associated with calculating a null distribution using permutations of data sets with large sample sizes. Slightly modifying our test statistic by centering $\mathbb{K}$, we can avoid the computational burden in obtaining the null distribution of $T$ \citep{Zhan2017}. Following the fast algorithm developed by \citet{Zhan2017}, we instead consider $T = \bmath{w}^T \mathbb{P} \mathbb{K} \mathbb{P} \bmath{w}$, where $\mathbb{P} = \mathbb{I}_n - \frac{1}{n} \bmath{1}\bmath{1}^T$ is the centering matrix. Because $\mathbb{K}$ is symmetric and positive semi-definite, there exists a matrix $\mathbb{K}^{\frac{1}{2}}$ such that $\mathbb{K} = \mathbb{K}^{\frac{1}{2}} (\mathbb{K}^{\frac{1}{2}})^T$. Thus, $T = \bmath{w}^T \mathbb{P} \mathbb{K} \mathbb{P} \bmath{w} = tr(\bmath{w}^T \mathbb{P} \mathbb{K} \mathbb{P} \bmath{w}) = tr(\mathbb{P}\mathbb{K}\mathbb{P}\bmath{w}\bmath{w}^T) = tr\{\mathbb{P} \mathbb{K}^{\frac{1}{2}} (\mathbb{K}^{\frac{1}{2}})^T \mathbb{P}\bmath{w}\bmath{w}^T\}$. 

This form of the statistic can then be leveraged with previously established distribution approximation methods. The analytical expressions of the first three moments of $T = tr(\mathbb{X}^T\mathbb{X}\mathbb{Y}^T\mathbb{Y})$ have been previously derived by \citet{Kazi-Aoual1995}. We may then utilize the fast techniques developed for RV-type statistics of the form $tr(\mathbb{P}\mathbb{A}\mathbb{A}^T\mathbb{P}\mathbb{B}\mathbb{B}^T)$, which approximate the null permutation distribution by moment matching to a Pearson type III distribution following \citet{Josse2008}. Although it is possible to approximate the empirical null distribution with the normal, log-normal, or Edgeworth distributions, the Pearson type III approximation is both accurate and more efficient than these competitors \citep{Josse2008}. Following the method described by \citet{Zhan2017}, we use the centered test statistic, analytical moments, and approximated Pearson type III density to obtain a p-value for our test statistic. In the case of $\rho(t) = 0.5|t|$, this faster algorithm has been shown to be about $6,000$ times faster than the standard permutation test with comparable power and type I error \citep{Zhan2017}. 

\begin{figure}[t]
\begin{tabular}{ccccc}
\rule{0pt}{0.01in}&$\epsilon_i \sim t_3$ & $\epsilon_i \sim \chi^2_1$ & $\epsilon_i \sim N(0,1)$ & $\epsilon_i \sim Cauchy(0,1)$ \\

 \raisebox{0.7in}{\rotatebox[origin=t]{90}{Linear}} & 
\subfloat{\includegraphics[width = 1.5in]{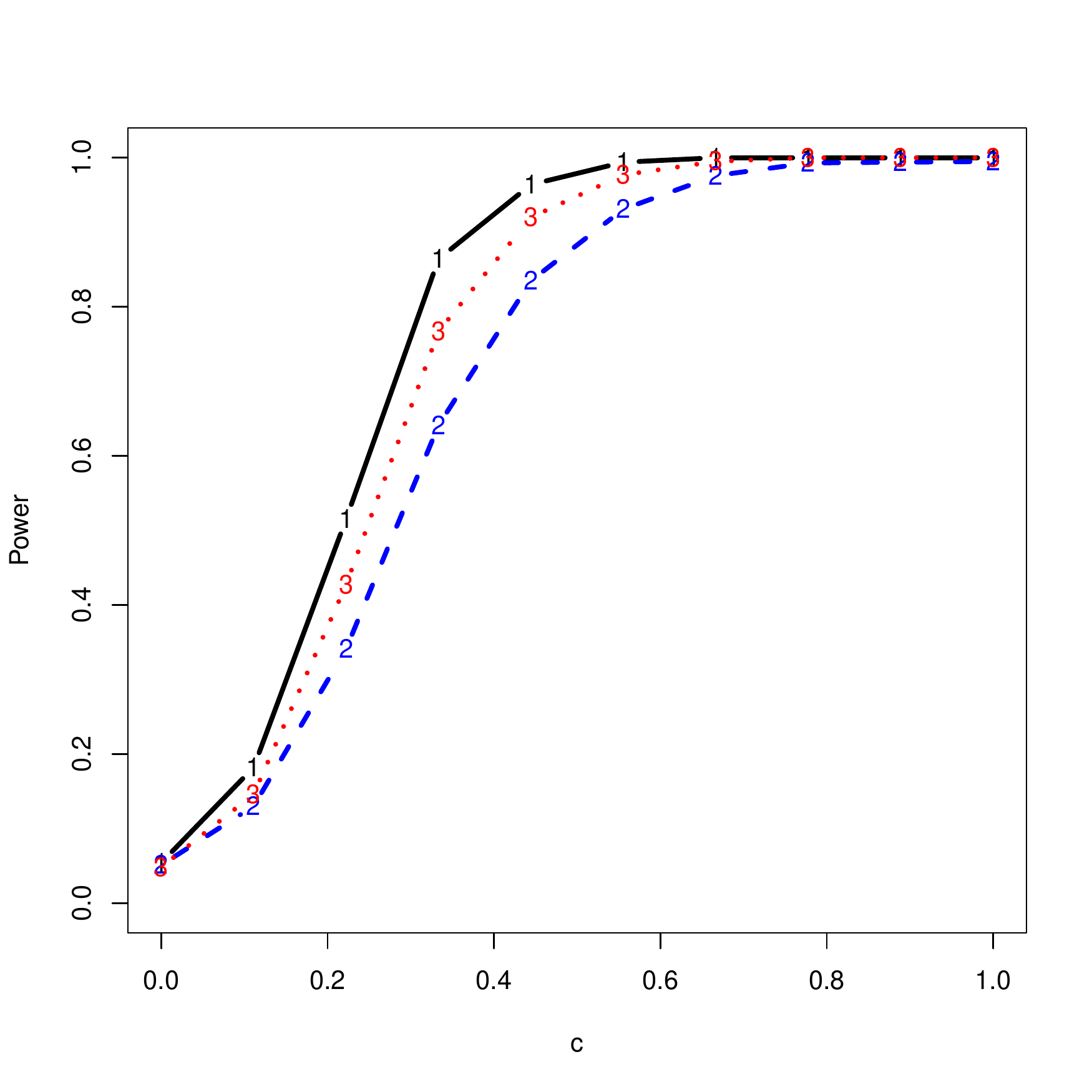}} &
\subfloat{\includegraphics[width = 1.5in]{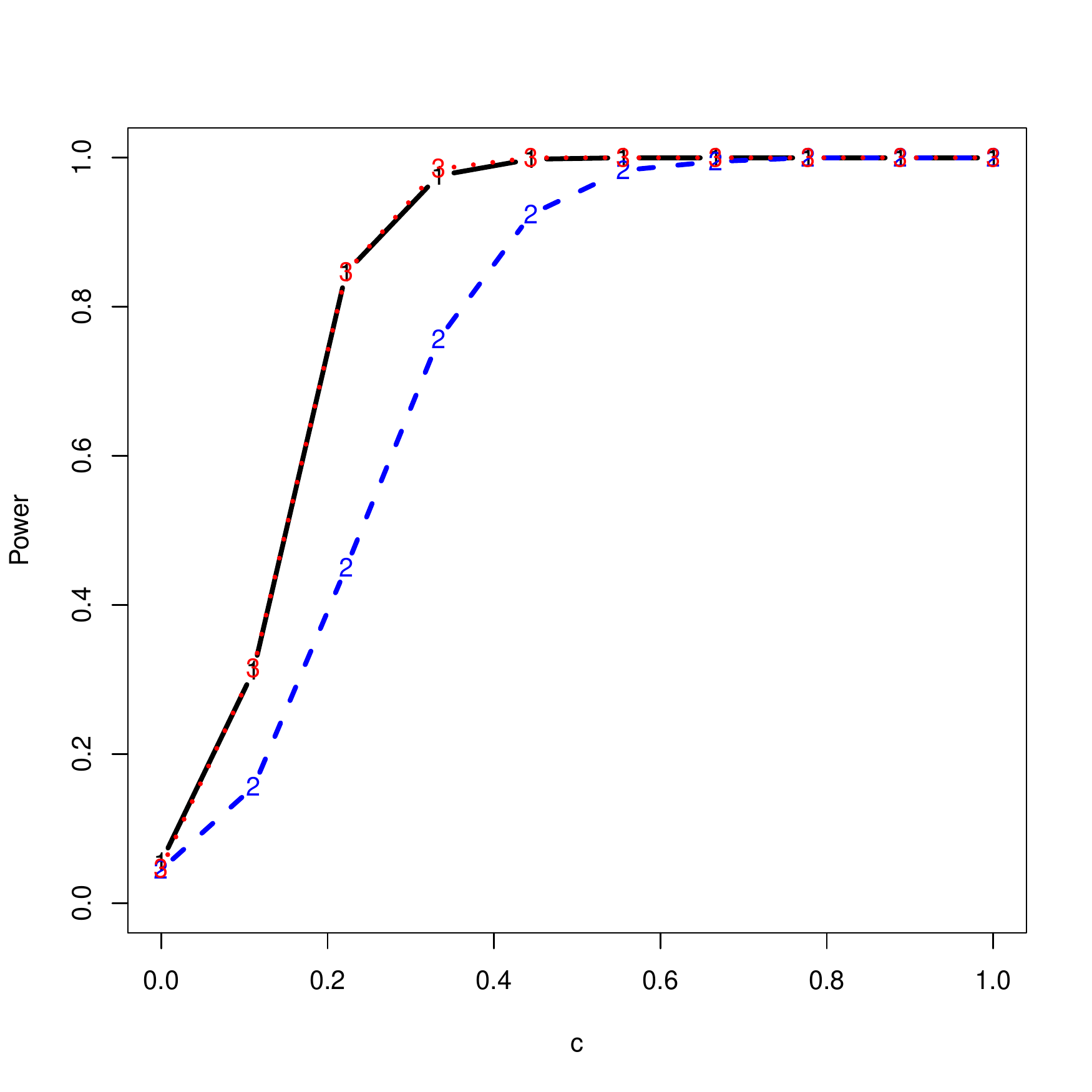}} &
\subfloat{\includegraphics[width = 1.5in]{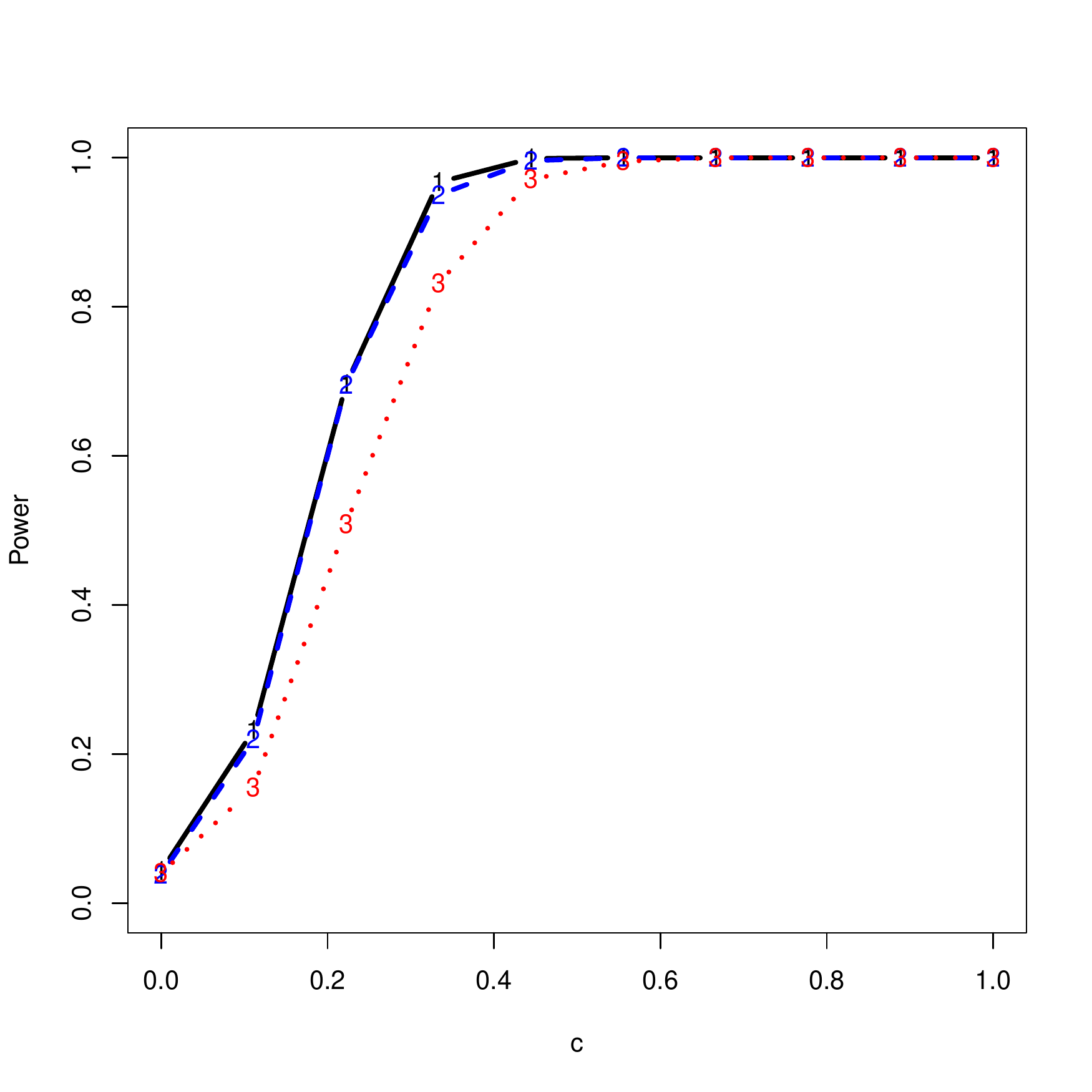}} &
\subfloat{\includegraphics[width = 1.5in]{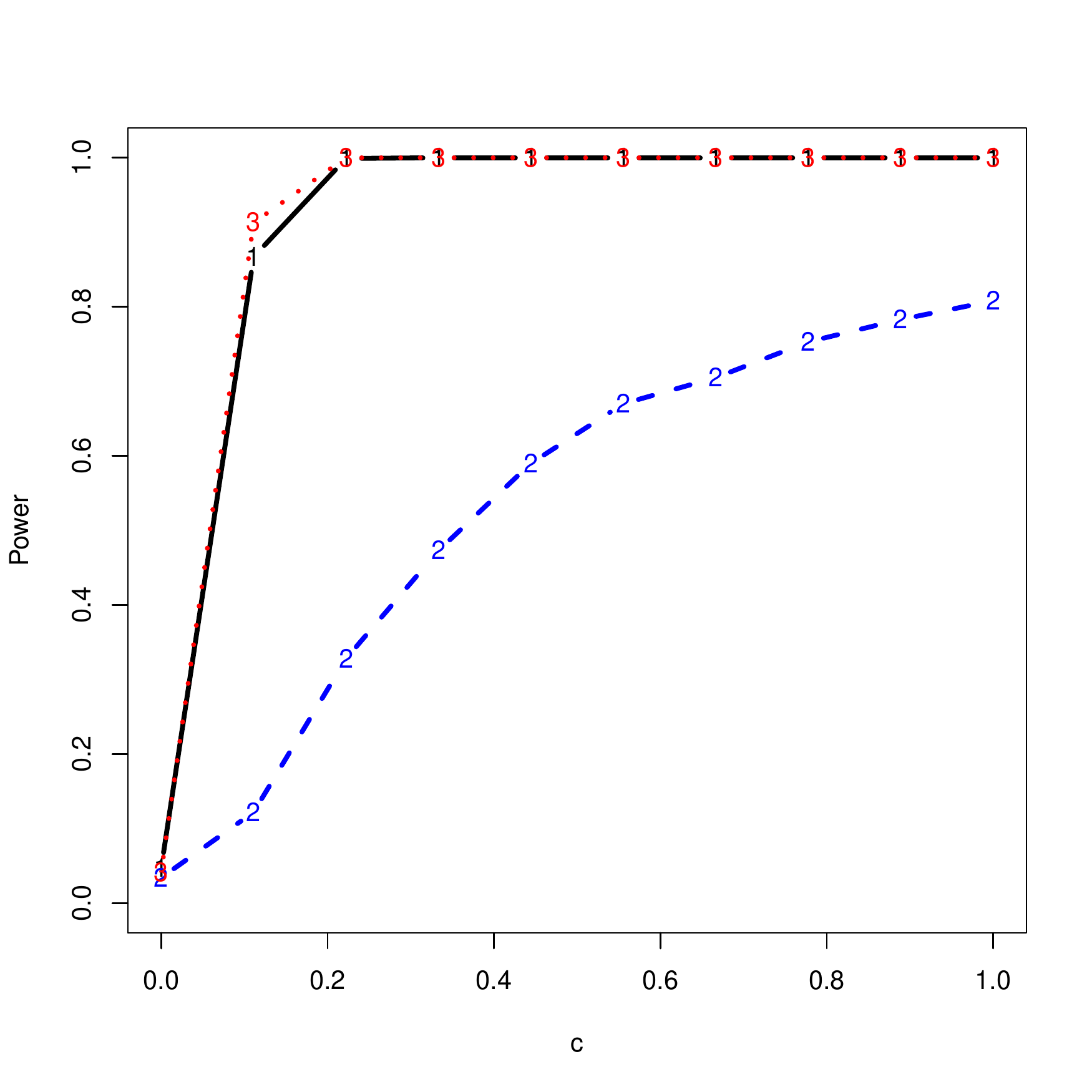}} \\

\raisebox{0.7in}{\rotatebox[origin=t]{90}{Nonlinear}} & 
\subfloat{\includegraphics[width = 1.5in]{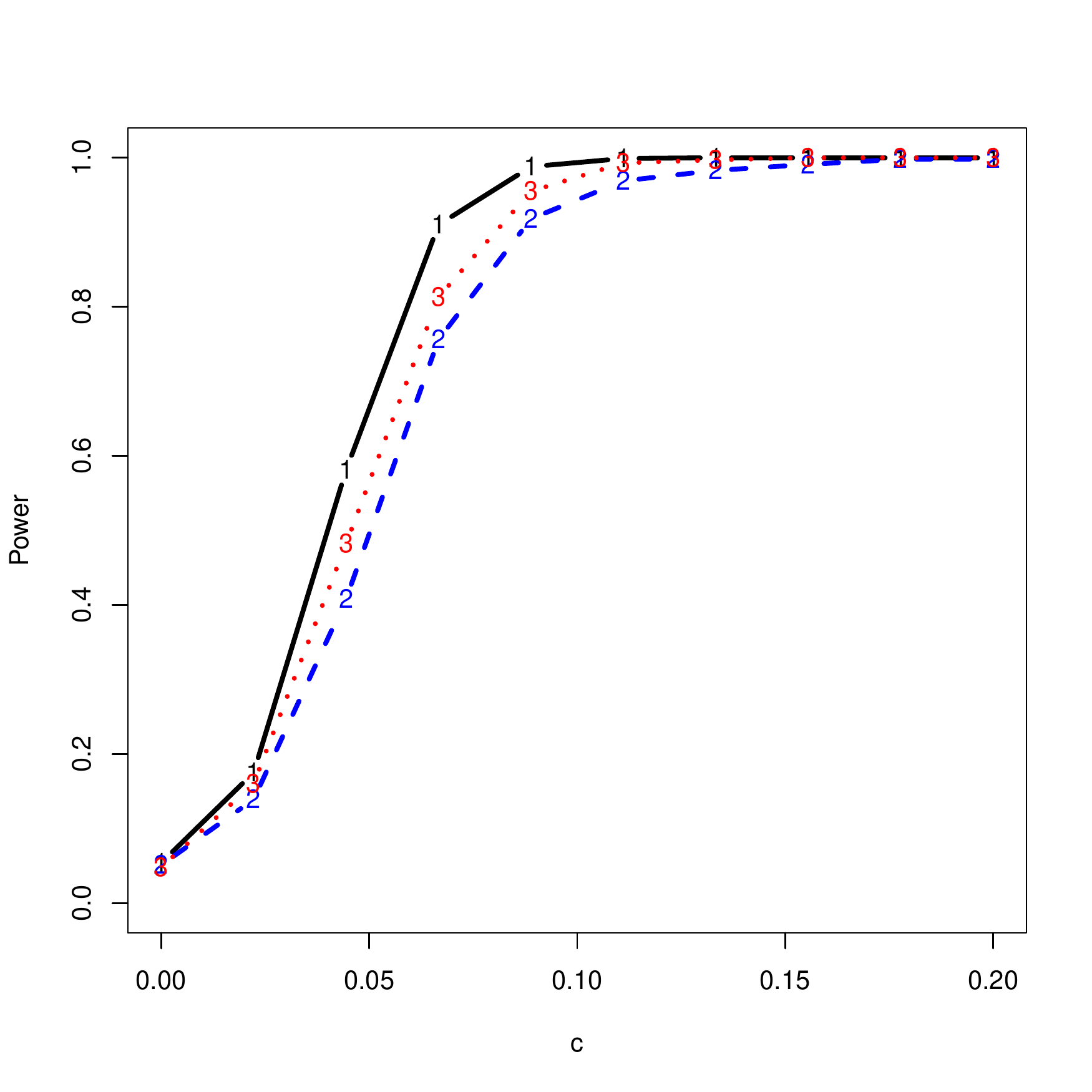}} &
\subfloat{\includegraphics[width = 1.5in]{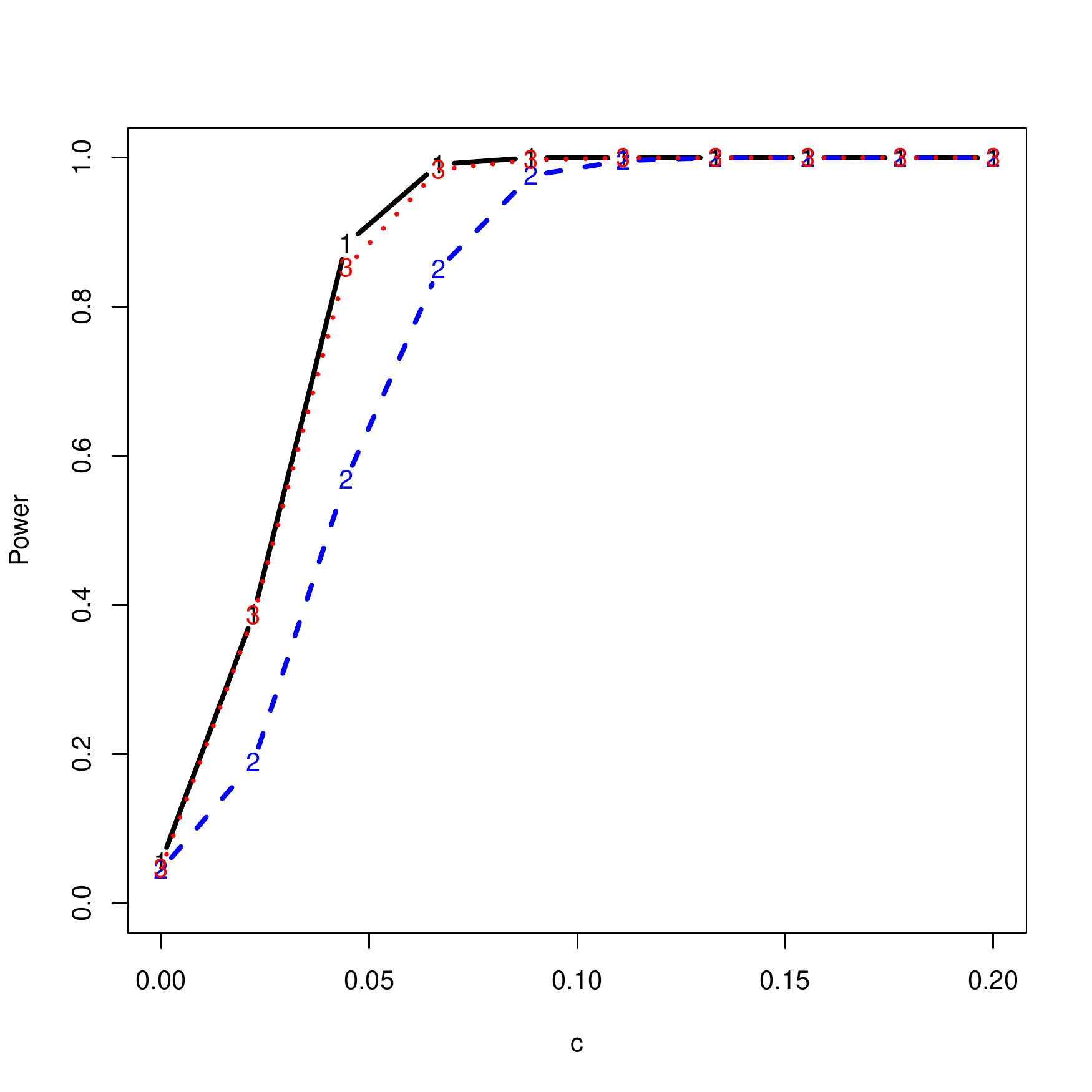}} &
\subfloat{\includegraphics[width = 1.5in]{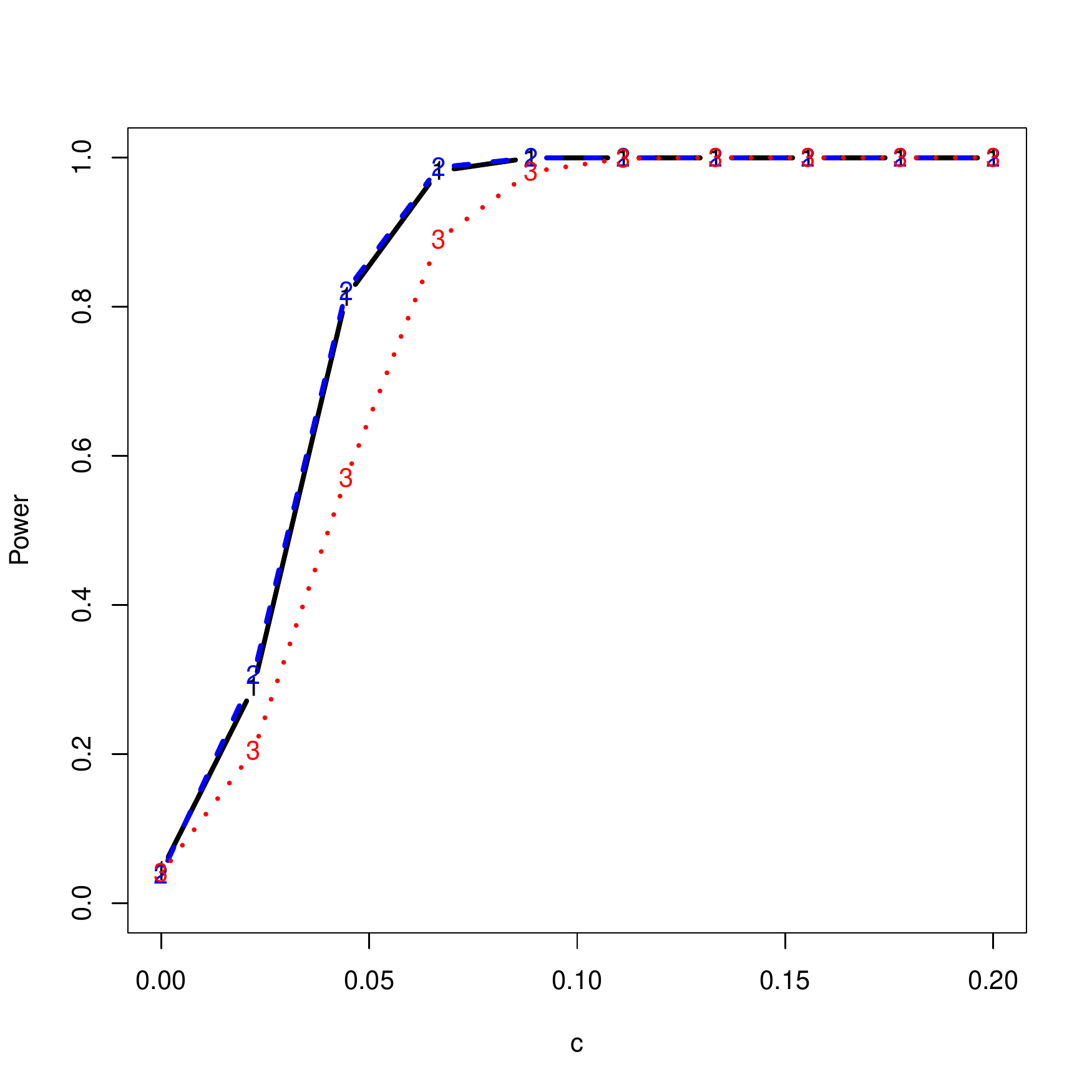}} &
\subfloat{\includegraphics[width = 1.5in]{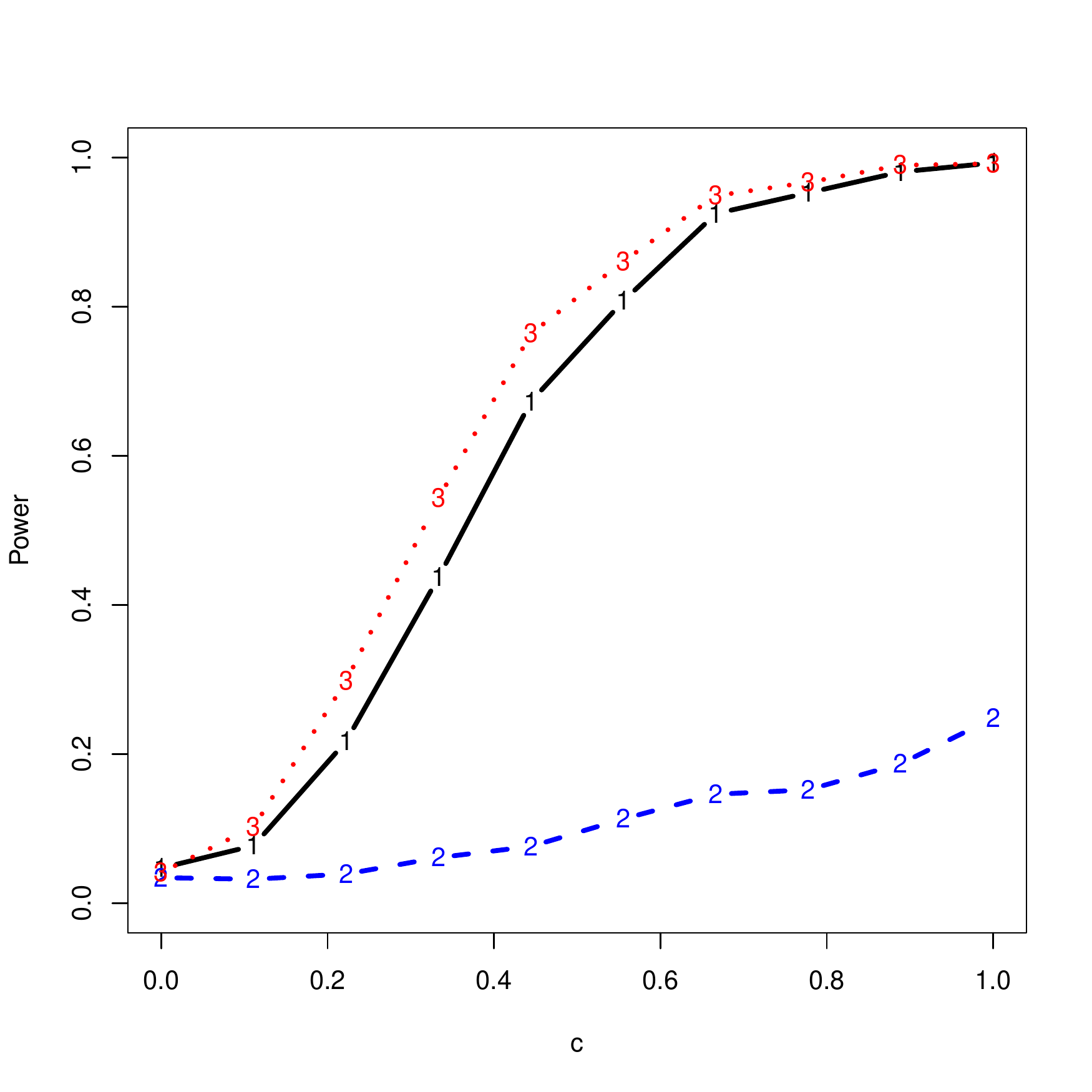}} 

\end{tabular}
\caption{\label{fig:powerIBS}Power simulations of RobKAT (black solid line), SKAT (blue dashed line), and LAD (red dotted line). The simulation uses IBS kernel, sample size $n=100$, and $1,000$ iterations at each setting of $c$. The top row corresponds to the linear case $h(\bmath{Z}_i) = \bmath{Z}_i^T \bmath{1}_5$, and the bottom row corresponds to the nonlinear case $h(\bmath{Z}_i) = 1 + \sum_{k=1}^p\bmath{Z}_{ik} + 2\sum_{k=2}^p \bmath{Z}_{i1}\bmath{Z}_{ik}$. Color figures appear in online versions of this article.}
\end{figure}

\section{Simulation Study}
\label{s:simulation}

We perform a simulation study to investigate the properties of our proposed method with the robust Huber $\psi$-function when the data have various error distributions and the genetic effect takes various forms. We then compare the performance of our proposed method against that of SKAT, which is not robust and assumes a normally distributed error distribution. We also include the special case of our robust method where $\rho(t) = 0.5|t|$, which corresponds to the QRKM method with $\tau = 0.5$. We refer to the QRKM $\rho(t) = 0.5|t|$ as least absolute deviation, LAD. 

\subsection{Simulation Study Design}
We generate data from the model \[Y_i = \bmath{X}_i^T \bmath{\beta} + c*h(\bmath{Z}_i) + \epsilon_i\] We generate elements of $\bmath{X}_i$ from $N(0, \mathbb{I}_5)$ and set $\bmath{\beta} = \bmath{1}_5$, and perform simulation studies with both the linear $h(\bmath{Z}_i) = \bmath{Z}_i^T \bmath{1}_5$ and the nonlinear $h(\bmath{Z}_i) = 1 + \sum_{k=1}^p\bmath{Z}_{ik} + 2\sum_{k=2}^p \bmath{Z}_{i1}\bmath{Z}_{ik}$, where $\bmath{Z}_i$ is a randomly sampled 9-SNP genotype of the \textit{SLC17A1} gene in the CATIE antibody study. We compare testing results from iid error terms, $\epsilon_i$, generated from $t_3$, $\chi^2_1$, $N(0,1)$, and $Cauchy(0, 1)$. Additionally, we compare the bimodal setting of a mixture of normal distributions, where $\epsilon_i = BW_0+  (1-B)W_{10} - 10(1-\theta)$ where $W_X \sim N(X,1)$, $B \sim Bern(\theta)$, and $\theta \in \{0.9, 0.7\}$. For all testing procedures, we use an IBS kernel and two sample sizes ($n= 100, 300$). We use the Huber function defined by $k=1.345$ for $95\%$ efficiency of least squares when the errors are normally distributed. 

\begin{figure}[h!]
\begin{tabular}{cccc}
\subfloat[$30\%$ mixture; n=100 \newline $h(\bmath{Z}_i)$ linear]{\includegraphics[width = 1.5in]{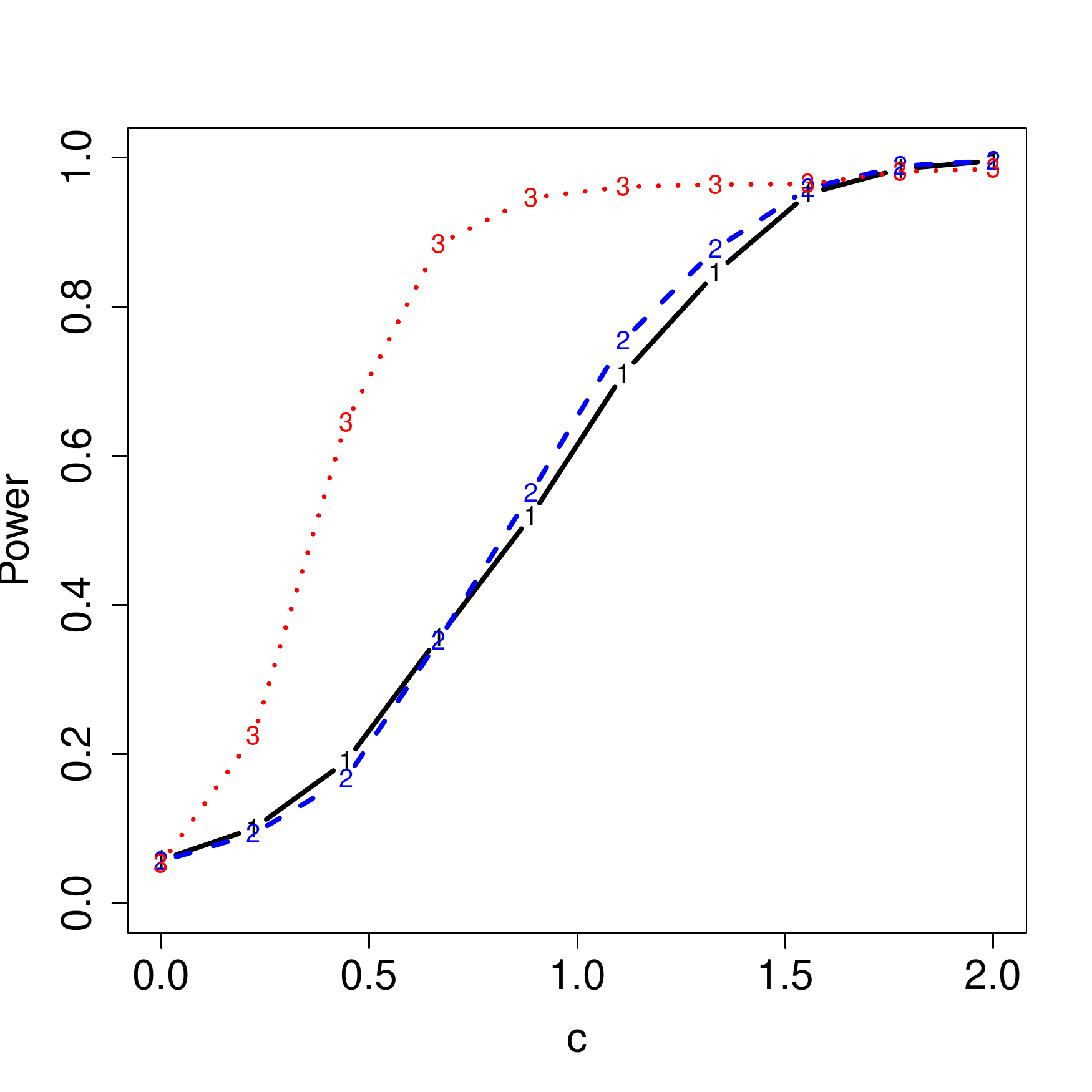}} &
\subfloat[$10\%$ mixture; n=100 \newline $h(\bmath{Z}_i)$ linear]{\includegraphics[width = 1.5in]{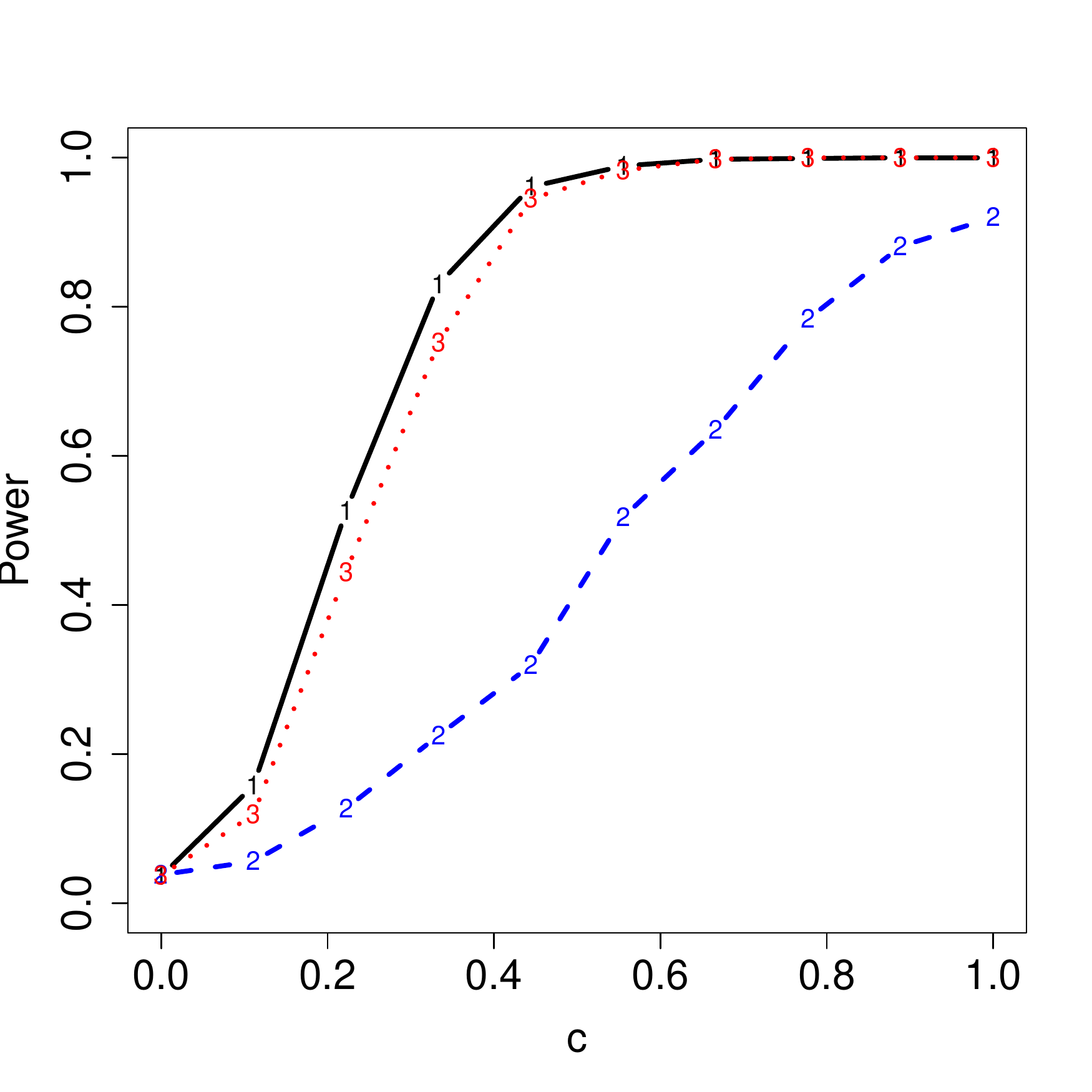}} &

\subfloat[$30\%$ mixture; n=100 \newline $h(\bmath{Z}_i)$ nonlinear]{\includegraphics[width = 1.5in]{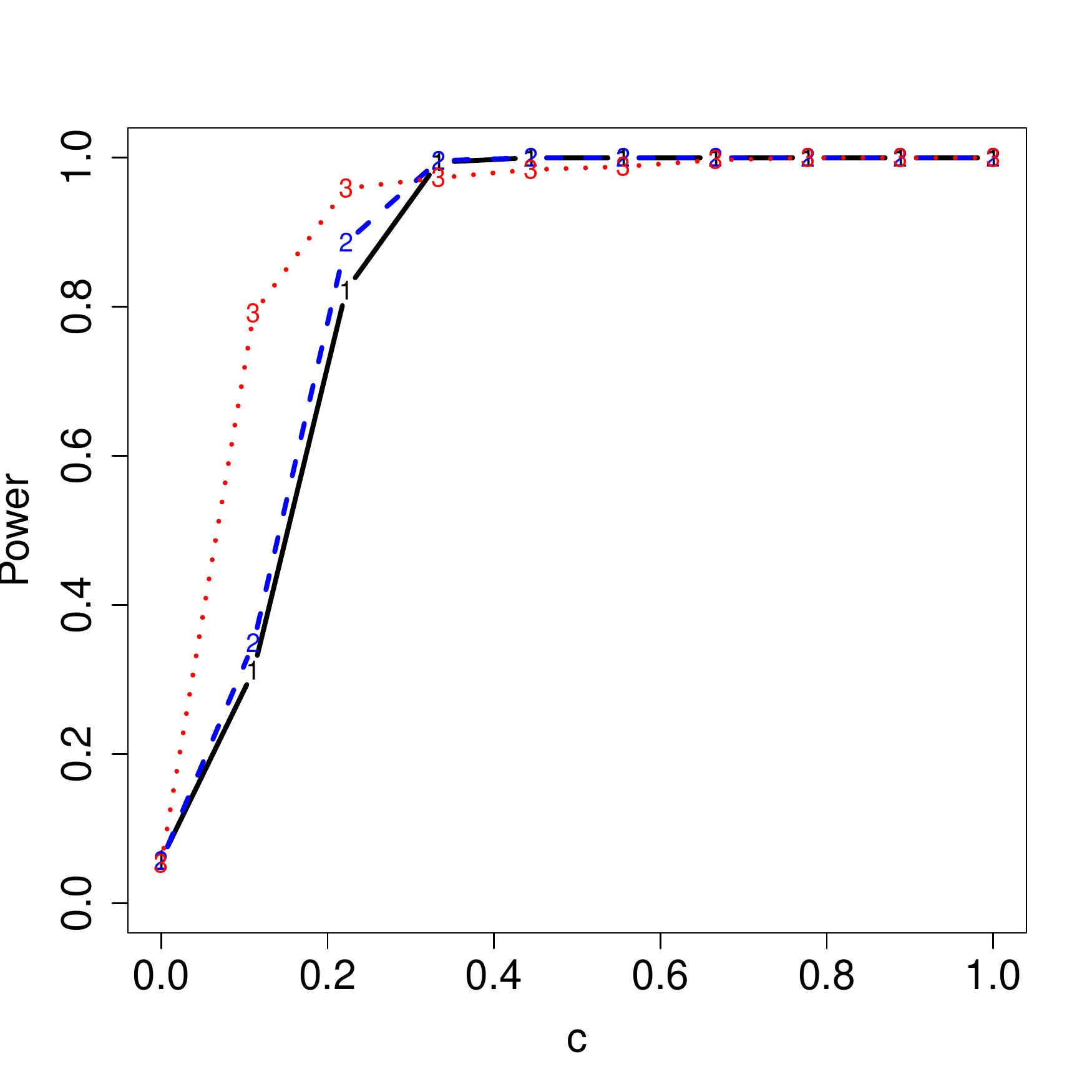}} &
\subfloat[$10\%$ mixture; n=100 \newline $h(\bmath{Z}_i)$ nonlinear]{\includegraphics[width = 1.5in]{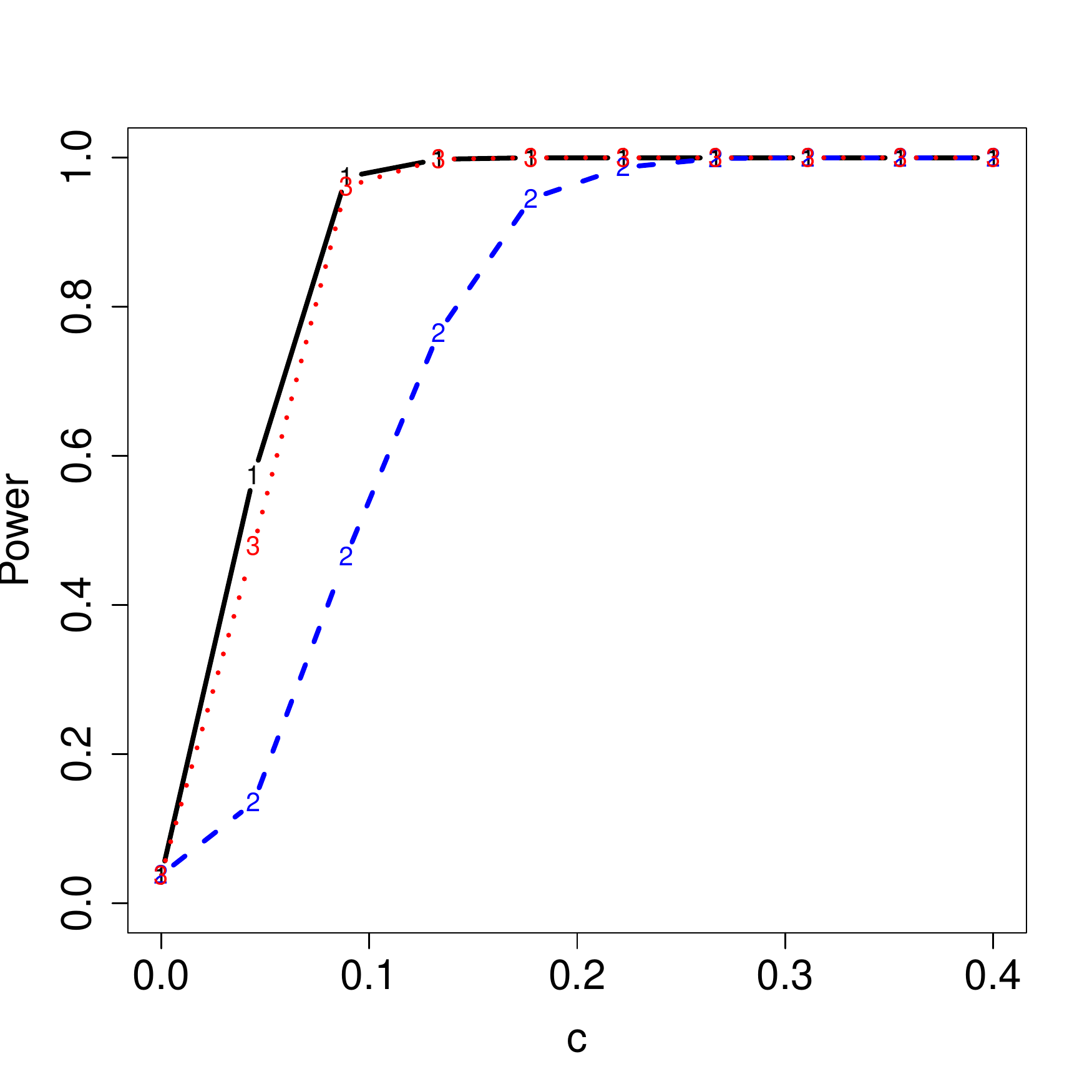}}
\end{tabular}
\caption{\label{fig:powerIBSmixture}Power simulations of RobKAT (black solid line), SKAT (green dotted line), and LAD (red dashedline). The simulation uses IBS kernel, sample size $n=100$, and $1,000$ iterations at each setting of $c$. Figures (a) and (b) show the linear case $h(\bmath{Z}_i) = Z_i^T 1_5$, and figures (c) and (d) display nonlinear case $h(\bmath{Z}_i) = 1 + \sum_{k=1}^p\bmath{Z}_{ik} + 2\sum_{k=2}^p \bmath{Z}_{i1}\bmath{Z}_{ik}$. Color figures appear in online versions of this article.}
\end{figure}

To examine Type I error, we consider significance levels $\alpha = 0.01$, $\alpha=0.05$, and $\alpha = 0.1$. We set $c=0$ for the null model under $H_0: h(\bmath{Z}_i) = 0$ and use $N=10,000$ Monte Carlo replications. As $c=0$, the type I errors are identical for both linear and nonlinear $h(\bmath{Z}_i)$ functions. To investigate the power of our testing procedure compared to SKAT, we test increasing values of $c$ and use $N=1,000$ Monte Carlo replications for each value. Figures~\ref{fig:powerIBS}-\ref{fig:powerIBSmixture} show the resulting power curves for tests with nominal level $0.05$.

\subsection{Simulation Results}
As seen in Table~\ref{tab:TypeI}, our testing method produces reasonable type I errors for each of the error distributions. Although SKAT produces similar type I errors for $t_3$, $\chi^2_1$, and $N(0,1)$, the type I errors with Cauchy distributed $\epsilon_i$ are notably deflated. Our testing method also produces reasonable type I errors across sample sizes for both the $10\%$ and $30\%$ normal mixtures, which is comparable to the SKAT testing method.

For standard normal and $t_3$ distributed errors, all tests displayed similar power. For the $\chi_1^2$ skewed errors, our robust method has similar power to LAD and a slight power advantage over SKAT. For Cauchy distributed errors, however, SKAT loses a substantial amount of power, while the Huber $\psi$-function method and LAD $\psi$-function perform reasonably well. For the $30\%$ normal mixture distribution, the Huber $\psi$-function performs similarly to SKAT, with the LAD $\psi$-function has a notably greater power than the other two methods. For the 10\% normal mixture distribution, the robust methods are more powerful than SKAT and similar to LAD. The association tests under the 30\% bimodal mixture overall is less powerful than the association test under the 10\% bimodal mixture. 

These trends are similar between the linear and nonlinear form of $h(\bmath{Z}_i)$. However, the nonlinear modeling of the genetic markers required smaller effect size for maximal power. 
The previously described trends for power are also consistent when the sample size increases to $n=300$ (See Web Appendix C).

\begin{table}[ht]
\centering
\begin{tabular}{ccrrr}
  \hline
  Error & $\alpha$ &  \multicolumn{3}{c}{{Loss Function ($\rho$)}}\\ [1pt]
  \hline
 && Huber & LAD & SKAT \\ 
  \hline
 \multirow{3}{*}{$t_3$} & $0.01$ & 0.0100 & 0.0083 & 0.0081 \\ 
  & $0.05$& 0.0456 & 0.0458 & 0.0466 \\ 
  & $0.10$& 0.0955 & 0.0943 & 0.0968 \\ \hline
  \multirow{3}{*}{$\chi^2_1$} & $0.01$ & 0.0105 & 0.0113 & 0.0080 \\ 
  &$0.05$ & 0.0532 & 0.0507 & 0.0475 \\ 
  & $0.10$& 0.0997 & 0.0999 & 0.0984 \\ \hline
   \multirow{3}{*}{$N(0,1)$} &$0.01$ & 0.0104 & 0.0089 & 0.0105 \\ 
  & $0.05$& 0.0473 & 0.0492 & 0.0455 \\ 
  & $0.10$& 0.0914 & 0.0977 & 0.0947 \\ \hline
   \multirow{3}{*}{$Cauchy$} &$0.01$ & 0.0120 & 0.0109 & 0.0032 \\ 
  & $0.05$& 0.0488 & 0.0509 & 0.0319 \\ 
  & $0.10$& 0.0973 & 0.1018 & 0.0841 \\ \hline
   \multirow{3}{*}{$10\%$ Mix} & $0.01$ & 0.0115 & 0.0094 & 0.0088 \\ 
  & $0.05$& 0.0484 & 0.0476 & 0.0494 \\ 
  & $0.10$& 0.0962 & 0.0944 & 0.1006 \\ \hline
   \multirow{3}{*}{$30\%$ Mix} & $0.01$ & 0.0108 & 0.0103 & 0.0096 \\ 
  & $0.05$& 0.0484 & 0.0487 & 0.0489 \\ 
  & $0.10$& 0.0996 & 0.0949 & 0.0986 \\ 
   \hline
\end{tabular}
\caption{\label{tab:TypeI}Simulation study for type I error of RobKAT (Huber loss), QRKM (LAD loss), and SKAT. The simulation includes IBS kernel, sample size $n=100$, and linear $h(\bmath{Z}_i) = 
\bmath{Z}_i^T \bmath{1}_5$. Estimates show proportion of $10,000$ iterations with p-value below $\alpha$ cutoff.}   
\end{table}

\section{Data Analysis}
\label{s:dataAnalysis}

Several genes in chromosome 6, including the Major Histocompatibility Complex (MHC), region have been found to have  an established connection to schizophrenia, and previous GWAS analyses have identified multiple SNPs in this region associated with schizophrenia \citep{Aberg2013, InternationalSchizophreniaConsortium2009, Shi2009}. As described in \citet{Lieberman2005}, the Clinical Antipsychotic Trials of Intervention Effectiveness (CATIE) study aimed to compare the effect of various atypical antipsychotics on unrelated patients' schizophrenia symptoms. DNA samples were donated from 51\% of the 1460 patients in the CATIE study, and genotyping was performed for 492,000 SNPs \citep{Sullivan2008}. 


Genetic variation in chromosome 6 is also thought to contribute to immune function and response to infectious diseases, such as neurotropic herpesviruses. This class of infectious agent has been linked to cognitive impairments, and in the case of Herpes Simplex Virus Type 1 have shown significant association with aspects of the decreased cognition found in schizophrenia disorders \citep{Yolken2011}. At baseline in the CATIE trial, participants provided a blood sample from which investigators measured Ig class antibodies to three neurotrophic herpesviruses: Herpes Simplex Virus type 1 (HSV1), Herpes Simplex Virus type 2 (HSV2), and Cytomegalovirus (CMV) \citep{Yolken2011}. 

We aim to apply our robust kernel machine methodology to determine genes near the MHC region with association to levels of three herpesvirus antibodies: HSV1, HSV2, and CMV. We investigate 12 genes on chromosome 6p22.1 including MHC region as was done previously in \cite{Maity2012} and \cite{Davenport2017}. For each gene, we perform our proposed kernel association test with the Huber $\psi$-function and three different kernels (linear, quadratic, IBS), controlling for age and sex. We use a Bonferroni multiple testing correction to an $\alpha=0.05$ significance level ($\alpha^* = 0.05 / (12*3*3) = 0.00046)$ as we are performing tests for 12 genes, 3 kernels, and 3 methods. For comparison, we also apply SKAT and LAD to analyze each of the 12 candidate genes with the same three kernel functions. The results for our analyses are summarized in Table \ref{tab:dataAnalysis3psi}.  

For the HSV1 and CMV antibody responses, no significant associations were detected for any gene with any loss function. For the HSV2 antibody response and each kernel, both SKAT and RobKAT detected significant association with the MHC region. However, our robust kernel association test was able to detect significant associations to the POM121L2 and SLC17A1 genes, while SKAT and LAD did not detect any significant associations. LAD detected a significant association with the HIST1H2BJ gene, while the other $\psi$-functions did not. 

It is sensible that both SKAT and the proposed test detected the MHC region, which has the most established association with both schizophrenia and immune function. However, the data for each antibody response are skewed (see Supplementary Materials for data distributions). As our previous simulations suggest improved power in non-normally distributed data sets, perhaps our model detected a signal in the SLC17A1 and POM121L2 genes whereas SKAT failed. The association between POM121L2 and schizophrenia has been previously noted by  \citet{Aberg2013}. The power described in the simulation study represents a probability, so it is not unreasonable that LAD detected HIST1H2BJ gene when our robust test did not.

\begin{table}[H]
	\centering
	\begin{adjustbox}{width=1\textwidth}
		\begin{tabular}{ccccccccccc}
			\hline
			\multicolumn{2}{c}{} & \multicolumn{3}{c}{HSV1} & \multicolumn{3}{c}{HSV2} &\multicolumn{3}{c}{CMV}\\
			\cmidrule(lr){3-5} \cmidrule(lr){6-8} \cmidrule(lr){9-11}
			gene&$\psi$& linear & quadratic & IBS & linear & quadratic & IBS & linear & quadratic & IBS \\ 
			\hline
			\multirow{3}{*}{BTN2A1} &SKAT & 0.07679 & 0.15524 & 0.10567 & 0.00176 & 0.00335 & 0.00494 & 0.00150 & 0.00759 & 0.00236 \\
			&huber & 0.00803 & 0.02631 & 0.01436 & 0.00541 & 0.00639 & 0.02336 & 0.00155 & 0.00863 & 0.00251 \\ 
            &LAD & 0.03475 & 0.10657 & 0.04764 & 0.00106 & 0.00150 & 0.00514 & 0.01024 & 0.03866 & 0.01116 \\
			\hline
			\multirow{3}{*}{BTN2A2}&SKAT & 0.05244 & 0.08935 & 0.05702 & 0.05251 & 0.06792 & 0.07835 & 0.00324 & 0.00763 & 0.00286 \\ 
			&huber & 0.01515 & 0.03424 & 0.01131 & 0.01443 & 0.02229 & 0.01920 & 0.00517 & 0.01266 & 0.00425 \\ 
            &LAD & 0.11892 & 0.21728 & 0.08933 & 0.03188 & 0.07616 & 0.03467 & 0.01244 & 0.01766 & 0.01149 \\
			\hline
			\multirow{3}{*}{BTN3A2}&SKAT & 0.03494 & 0.02282 & 0.01762 & 0.07271 & 0.06011 & 0.08863 & 0.04910 & 0.03126 & 0.07955 \\ 
			&huber & 0.01431 & 0.00673 & 0.01121 & 0.15886 & 0.19427 & 0.19263 & 0.07666 & 0.04850 & 0.11918 \\ 
            &LAD & 0.00384 & 0.00212 & 0.00339 & 0.84444 & 0.73582 & 0.90418 & 0.31518 & 0.23253 & 0.44299 \\
			\hline
			\multirow{3}{*}{HIST1H2AG}&SKAT & 0.92213 & 0.93267 & 0.70116 & 0.19853 & 0.23608 & 0.25445 & 0.43686 & 0.66758 & 0.21429 \\ 
			&huber & 0.72157 & 0.81800 & 0.85490 & 0.06201 & 0.12390 & 0.03250 & 0.38483 & 0.55682 & 0.25441 \\ 
            &LAD & 0.48233 & 0.38472 & 0.80218 & 0.00241 & 0.00306 & 0.00407 & 0.56707 & 0.73559 & 0.45521 \\
			\hline
			\multirow{3}{*}{\textbf{HIST1H2BJ}}&SKAT & 0.79122 & 0.58380 & 0.70037 & 0.01310 & 0.01899 & 0.01649 & 0.18609 & 0.30461 & 0.11541 \\ 
			&huber & 1.00000 & 1.00000 & 1.00000 & 0.00121 & 0.00236 & 0.00129 & 0.16770 & 0.25322 & 0.13759 \\ 
            &LAD & 1.00000 & 1.00000 & 1.00000 & \textbf{0.00028} & \textbf{0.00018} & 0.00098& 0.27022 & 0.21404 & 0.26863 \\
			\hline
			\multirow{3}{*}{\textbf{MHC}}&SKAT & 0.04048 & 0.08628 & 0.04483 & \textbf{0.00010} & \textbf{0.00016 }& \textbf{0.00013} & 0.10080 & 0.14007 & 0.10371 \\ 
			&huber & 0.00877 & 0.01932 & 0.00660 & \textbf{2.2314e-06}& \textbf{1.7107e-06} & \textbf{2.4198e-06 }& 0.05149 & 0.06349 & 0.05627 \\ 
            &LAD & 0.03487 & 0.02926 & 0.02175 & 0.00419 & 0.00159 & 0.00180 & 0.07406 & 0.07532 & 0.11273 \\
			\hline
			\multirow{3}{*}{NOTCH4}&SKAT & 0.04606 & 0.04072 & 0.06212 & 0.26256 & 0.40066 & 0.28563 & 0.60714 & 0.93015 & 0.64687 \\ 
			&huber & 0.03791 & 0.03982 & 0.05217 & 0.06431 & 0.21994 & 0.08140 & 0.43858 & 0.79330 & 0.48190 \\ 
            &LAD & 0.04826 & 0.05320 & 0.07832 & 0.16485 & 0.44452 & 0.25259 & 0.28569 & 0.53319 & 0.33174 \\
			\hline
			\multirow{3}{*}{\textbf{POM121L2}}&SKAT & 0.26827 & 0.36745 & 0.20218 & 0.00073 & 0.00146 & 0.00076 & 0.10933 & 0.16350 & 0.07037 \\ 
			&huber & 0.21433 & 0.29563 & 0.15850 & \textbf{0.00023} & \textbf{0.00042} & \textbf{0.00021 }& 0.09777 & 0.13930 & 0.07341 \\ 
            &LAD & 0.03044 & 0.05003 & 0.02184 & 0.00725 & 0.01584 & 0.00337 & 0.01474 & 0.01903 & 0.00989 \\
			\hline
			\multirow{3}{*}{PRSS16}&SKAT & 0.41067 & 0.41067 & 0.41067 & 0.65578 & 0.65578 & 0.65578 & 0.14173 & 0.14173 & 0.14173 \\ 
			&huber & 0.24503 & 0.24503 & 0.24226 & 0.32980 & 0.32980 & 0.33131 & 0.07214 & 0.07214 & 0.07305 \\ 
            &LAD & 0.09568 & 0.10038 & 0.10510 & 0.26890 & 0.26890 & 0.26107 & 0.03620 & 0.03620 & 0.03802 \\
			\hline
			\multirow{3}{*}{\textbf{SLC17A1}}&SKAT & 0.19398 & 0.17851 & 0.32453 & 0.00122 & 0.00396 & 0.00250 & 0.00529 & 0.01382 & 0.00961 \\ 
			&huber & 0.08473 & 0.09419 & 0.14602 & \textbf{0.00020} & 0.00070 & 0.00062 & 0.00492 & 0.01555 & 0.00913 \\ 
            &LAD & 0.24536 & 0.33055 & 0.39316 & 0.00400 & 0.00265 & 0.00420 & 0.03292 & 0.07575 & 0.04848 \\
			\hline
			\multirow{3}{*}{SLC17A3}&SKAT & 0.57656 & 0.40083 & 0.66907 & 0.02843 & 0.07313 & 0.05747 & 0.09016 & 0.05664 & 0.06619 \\ 
			&huber & 0.48986 & 0.39620 & 0.43752 & 0.01814 & 0.05846 & 0.03350 & 0.10804 & 0.06930 & 0.10833 \\ 
            &LAD & 0.97778 & 0.78567 & 0.66117 & 0.05555 & 0.17217 & 0.07699 & 0.16769 & 0.14177 & 0.13404 \\
			\hline
			\multirow{3}{*}{ZNF184}&SKAT & 0.58984 & 0.64917 & 0.61829 & 0.29280 & 0.19342 & 0.33406 & 0.32298 & 0.32206 & 0.44491 \\ 
			&huber & 0.46517 & 0.51445 & 0.63247 & 0.42303 & 0.39284 & 0.61290 & 0.30613 & 0.30321 & 0.39351 \\ 
            &LAD & 0.41699 & 0.46274 & 0.54990 & 0.58686 & 0.69201 & 0.61074 & 0.10163 & 0.10491 & 0.15728 \\ 
			\hline
		\end{tabular}
	\end{adjustbox}
	
	\caption{\label{tab:dataAnalysis3psi} Data analysis of CATIE data using RobKAT (huber loss), QRKM (LAD loss), and SKAT. The significance threshold was determined via a Bonferroni correction based on 12 gene analyses, 3 kernels, and 3 methods ($\alpha = 0.05 / (12*3*3) =0.00046$). Significant results are shown in bold.}
\end{table}

\section{Discussion}
\label{s:discuss}

We have developed the methodology of a general and robust kernel association test for the model $Y_i = \bmath{X}_i^T \bmath{\beta} + h(\bmath{Z}_i) + \epsilon_i$ with the score-type test statistic $T = \bmath{w}^t \mathbb{K} \bmath{w}$. Our kernel association test is general, and many previously established kernel association tests such as SKAT and QRKM are special cases when specific loss functions are chosen. RobKAT is flexible in that it places few assumptions on both the choice of kernel and the choice of loss function. The test also does not place any assumptions, including existence of moments, on the response distribution. We then implemented RobKAT with fast permutation testing techniques, which also provides the null distribution of our test statistic without assuming a response distribution. This permutation testing approach is key, as the LSKM Chi-squared mixture null distribution with our statistic decreases power and fails to control for type I error in small and moderate sample sizes (See Web Appendix F for simulation evidence).

We ran a simulation study with an IBS kernel, three different loss functions (Huber, LAD, SKAT), and various response distributions ($t_3, \chi^2_1, N(0,1), Cauchy(0,1)$) to investigate the performance of our robust kernel association test. In all distributional settings, the type I error was controlled and the robust tests showed similar or greater power than SKAT. This pattern also held in a simulation study bimodal distributed responses. Furthermore, as no single robust loss function was universally superior across response distributions, there is a demonstrated benefit of allowing a variety of loss functions, which is a feature of RobKAT.  

We were also able to apply our methodology to the CATIE data set to test for associations between levels of neurotropic herpesviruses and genes located in the MHC region of chromosome 6. Although the antibody level response was non-normal, our robust kernel association test was able to detect significant associations between HSV2 and the MHC region and the HIST1H2BJ, POM121L2, and SLC17A1 genes. Among these genes, SKAT only detected association between HSV2 levels and the MHC region, which is unsurprising given the observed trend from simulation that SKAT is less powerful than our robust test when the response is non-normal. 

Simulation studies for linear and quadratic kernels were also performed and produced similar power and type I error trends among the various loss functions. For both the linear and nonlinear forms of $h(\bmath{Z}_i)$, the tests utilizing different kernels had similar power with the quadratic kernel producing the greatest power (See Web Appendices A-B for full simulation results).

Simulation studies for two additional $\psi$-functions, Hampel and Bisquare, were also performed (See Appendix). The Hampel family of functions is defined by 

\[\psi(x) = \begin{cases} x &  |x|\leq a \\ a \text{ sign}(x) &  a < |x| \leq b \\ a \text{ sign}(x)\frac{r-|x|}{r-b} & b < |x| \leq r \\ 0 & r < |x| \end{cases}\]
and setting $a=1.353, b=3.157, r=7.216$ in our simulations for $95\%$ efficiency \citep{Hampel1986}. Tukey's Bisquare family of functions is defined by $\psi(x) = x\bigg(1-(\frac{x}{k})^2\bigg)^2 I_{\{|x|\leq k\}}$ and setting $k = 4.685$ for $95\%$ efficiency in our simulations \citep{Maronna2006}. These two $\psi$-functions are called redescending functions, meaning that there exists a rejection point $c < \infty$ such that $\psi(x) = 0$ for all $|x| \geq c$ \citep{Huber1981}. 

Although these redescending functions violate our assumptions on $\psi$, redescending functions may be attractive in data sets with extreme outliers in the distribution's tails, as the outliers would likely receive no weight in model fitting. Some researchers such as \citet{Huber1981} warn against the use of these redescending functions, arguing that the risks from sensitivity to incorrect $\psi$ scale parameter specification do not outweigh the slight improvement in asymptotic variance. For these reasons, simulation results that include the Hampel and Bisquare $\psi$-functions are only included in the Supplementary Web Materials.

For the Cauchy and 30\% mixture distributions, the redescending $\psi$-functions had slightly more power than with the other $\psi$-functions, likely due to the heavy tails and outliers in these distributions. For all other distributions, the power and type I error control were similar to that of the Huber $\psi$-function. With increasing effect sizes in our simulation study, we observed an eventual decreasing power trend for the redescending $\psi$-functions. With large effect size, we suspect that the plato in $\rho$-function inherent from the redescending $\psi$ fails to detect strong signals.

Our algorithm is also computationally efficient through the utilization of fast kernel machine testing methods developed by \citet{Zhan2017}. Thus, not only is the RobKAT test general and robust, but our robust kernel association test is also scalable to moderate and large amounts of SNPs. For example, the average time to perform one RobKAT test with sample size 300, IBS kernel, and linear $h(\bmath{Z}_i)$ is 0.17 seconds.
%
%
Additionally, we have produced a user-friendly R package containing the RobKAT software. Thus, our robust kernel association test methodology is a robust generalization with fewer distributional and inferential restrictions than previously established methods, while remaining computationally efficient and user friendly.









\backmatter


\section*{Acknowledgements}
Data used in the preparation of this article were obtained from the limited access datasets distributed from the NIH-supported ``Clinical Antipsychotic Trials of Intervention Effectiveness in Alzheimer's Disease" (CATIE-AD). This is a multisite, clinical trial of persons with Alzheimer's Disease comparing the effectiveness of randomly assigned medication treatment. The study was supported by NIMH Contract \#N01MH90001 to the University of North Carolina at Chapel Hill. The ClinicalTrials.gov identifier is NCT00015548. This work was partially supported by National Institutes of Health grant P01CA142538 (to JYT) and National Institutes of Health training grant T32GM081057, Biostatistics Training in the Omics Era (to KM).



\section*{Supplementary Materials}

Supplementary simulation and data analysis results are available with
this paper at the Biometrics website on Wiley Online
Library.\vspace*{-8pt}


%

\begin{thebibliography}{}

\bibitem[\protect\citeauthoryear{Aberg, Liu, Bukszar, McClay, Khachane,
  Andreassen, Blackwood, Corvin, Djurovic, Gurling, Ophoff, Pato, Pato, Riley,
  Webb, Kendler, O'Donovan, Craddock, Kirov, Owen, Rujescu, StClair, Werge, and
  etAl}{Aberg et~al.}{2013}]{Aberg2013}
Aberg, K., Liu, Y., Bukszar, J., McClay, J., Khachane, A., Andreassen, O.,
  Blackwood, D., Corvin, A., Djurovic, S., Gurling, H., Ophoff, R., Pato, C.,
  Pato, M., Riley, B., Webb, T., Kendler, K., O'Donovan, M., Craddock, N.,
  Kirov, G., Owen, M., Rujescu, D., StClair, D., Werge, T., and etAl (2013).
\newblock A comprehensive family-based replication study of schizophrenia
  genes.
\newblock{\em JAMA Psychiatry} {\bf 70,} 573.

\bibitem[\protect\citeauthoryear{Davenport, Maity, Sullivan, and
  Tzeng}{Davenport et~al.}{2018}]{Davenport2017}
Davenport, C.~A., Maity, A., Sullivan, P.~F., and Tzeng, J.-Y. (2018).
\newblock{A Powerful Test for SNP Effects on Multivariate Binary Outcomes
  using Kernel Machine Regression}.
\newblock{\em Statistics in Biosciences} {\bf 10,} 117--138.

\bibitem[\protect\citeauthoryear{Gauderman, Murcray, Gilliland, and
  Conti}{Gauderman et~al.}{2007}]{Gauderman2007}
Gauderman, W.~J., Murcray, C., Gilliland, F., and Conti, D.~V. (2007).
\newblock{Testing association between disease and multiple SNPs in a candidate
  gene}.
\newblock{\em Genetic Epidemiology} {\bf 31,} 383--395.

\bibitem[\protect\citeauthoryear{Hampel}{Hampel}{1986}]{Hampel1986}
Hampel, F.~R. (1986).
\newblock{\em {Robust statistics : the approach based on influence
  functions}}.
\newblock Wiley, New York.

\bibitem[\protect\citeauthoryear{Huber}{Huber}{1977}]{PeterHuber1977}
Huber, P.~J. (1977).
\newblock{\em {Robust statistical procedures}}.
\newblock Society for Industrial and Applied Mathematics, Philadelphia, PA, 1
  edition.

\bibitem[\protect\citeauthoryear{Huber}{Huber}{1981}]{Huber1981}
Huber, P.~J. (1981).
\newblock{\em {Robust statistics}}.
\newblock Wiley, Sommerset, NJ, 1 edition.

\bibitem[\protect\citeauthoryear{{International Schizophrenia Consortium},
  Purcell, Wray, Stone, Visscher, O'Donovan, Sullivan, and
  Sklar}{{International Schizophrenia Consortium}
  et~al.}{2009}]{InternationalSchizophreniaConsortium2009}
{International Schizophrenia Consortium}, I.~S., Purcell, S.~M., Wray, N.~R.,
  Stone, J.~L., Visscher, P.~M., O'Donovan, M.~C., Sullivan, P.~F., and Sklar,
  P. (2009).
\newblock{Common polygenic variation contributes to risk of schizophrenia and
  bipolar disorder.}
\newblock{\em Nature} {\bf 460,} 748--52.

\bibitem[\protect\citeauthoryear{Josse, Pag{\`{e}}s, and Husson}{Josse
  et~al.}{2008}]{Josse2008}
Josse, J., Pag{\`{e}}s, J., and Husson, F. (2008).
\newblock{Testing the significance of the RV coefficient}.
\newblock{\em Computational Statistics and Data Analysis} .

\bibitem[\protect\citeauthoryear{Kazi-Aoual, Hitier, Sabatier, and
  Lebreton}{Kazi-Aoual et~al.}{1995}]{Kazi-Aoual1995}
Kazi-Aoual, F., Hitier, S., Sabatier, R., and Lebreton, J.-D. (1995).
\newblock{Refined approximations to permutation tests for multivariate
  inference}.
\newblock{\em Computational Statistics {\&} Data Analysis} {\bf 20,} 643--656.

\bibitem[\protect\citeauthoryear{Kong, Maity, Hsu, and Tzeng}{Kong
  et~al.}{2016}]{Kong2016}
Kong, D., Maity, A., Hsu, F.-C., and Tzeng, J.-Y. (2016).
\newblock{Testing and estimation in marker-set association study using
  semiparametric quantile regression kernel machine}.
\newblock{\em Biometrics} {\bf 72,} 364--371.

\bibitem[\protect\citeauthoryear{Kwee, Liu, Lin, Ghosh, and Epstein}{Kwee
  et~al.}{2008}]{Kwee2008}
Kwee, L.~C., Liu, D., Lin, X., Ghosh, D., and Epstein, M.~P. (2008).
\newblock{A Powerful and Flexible Multilocus Association Test for Quantitative
  Traits}.
\newblock{\em The American Journal of Human Genetics} {\bf 82,} 386--397.

\bibitem[\protect\citeauthoryear{Li and Leal}{Li and Leal}{2008}]{Li2008}
Li, B. and Leal, S.~M. (2008).
\newblock{Methods for Detecting Associations with Rare Variants for Common
  Diseases: Application to Analysis of Sequence Data}.
\newblock{\em The American Journal of Human Genetics} {\bf 83,} 311--321.

\bibitem[\protect\citeauthoryear{Lieberman, Stroup, McEvoy, Swartz, Rosenheck,
  Perkins, Keefe, Davis, Davis, Lebowitz, Severe, and Hsiao}{Lieberman
  et~al.}{2005}]{Lieberman2005}
Lieberman, J.~A., Stroup, T.~S., McEvoy, J.~P., Swartz, M.~S., Rosenheck,
  R.~A., Perkins, D.~O., Keefe, R.~S., Davis, S.~M., Davis, C.~E., Lebowitz,
  B.~D., Severe, J., and Hsiao, J.~K. (2005).
\newblock{Effectiveness of Antipsychotic Drugs in Patients with Chronic
  Schizophrenia}.
\newblock{\em New England Journal of Medicine} {\bf 353,} 1209--1223.

\bibitem[\protect\citeauthoryear{Liu, Lin, and Ghosh}{Liu
  et~al.}{2007}]{Liu2007}
Liu, D., Lin, X., and Ghosh, D. (2007).
\newblock{Semiparametric Regression of Multidimensional Genetic Pathway Data:
  Least-Squares Kernel Machines and Linear Mixed Models}.
\newblock{\em Biometrics} {\bf 63,} 1079--1088.

\bibitem[\protect\citeauthoryear{Madsen and Browning}{Madsen and
  Browning}{2009}]{Madsen2009}
Madsen, B.~E. and Browning, S.~R. (2009).
\newblock{A Groupwise Association Test for Rare Mutations Using a Weighted Sum
  Statistic}.
\newblock{\em PLoS Genetics} {\bf 5,} e1000384.

\bibitem[\protect\citeauthoryear{Maity, Sullivan, and Tzeng}{Maity
  et~al.}{2012}]{Maity2012}
Maity, A., Sullivan, P.~F., and Tzeng, J.-Y. (2012).
\newblock{Multivariate phenotype association analysis by marker-set kernel
  machine regression.}
\newblock{\em Genetic epidemiology} {\bf 36,} 686--95.

\bibitem[\protect\citeauthoryear{Maronna, Martin, and Yohai}{Maronna
  et~al.}{2006}]{Maronna2006}
Maronna, R.~A., Martin, R.~D., and Yohai, V.~J. (2006).
\newblock{\em {Robust Statistics}}.
\newblock Wiley Series in Probability and Statistics. John Wiley {\&} Sons,
  Ltd, Chichester, UK.

\bibitem[\protect\citeauthoryear{Schrader and Hettmansperger}{Schrader and
  Hettmansperger}{1980}]{Schrader1980}
Schrader, R.~M. and Hettmansperger, T.~P. (1980).
\newblock{Robust Analysis of Variance Based Upon a Likelihood Ratio
  Criterion}.
\newblock{\em Biometrika} {\bf 67,} 93.

\bibitem[\protect\citeauthoryear{Shawe-Taylor and Cristianini}{Shawe-Taylor and
  Cristianini}{2004}]{Shawe2004}
Shawe-Taylor, J. and Cristianini, N. (2004).
\newblock{\em {Kernel methods for pattern analysis}}.
\newblock Cambridge University Press, 1 edition.

\bibitem[\protect\citeauthoryear{Shi, Levinson, Duan, Sanders, Zheng, Pe'er,
  Dudbridge, Holmans, Whittemore, Mowry, Olincy, Amin, Cloninger, Silverman,
  Buccola, Byerley, Black, Crowe, Oksenberg, Mirel, Kendler, Freedman, and
  Gejman}{Shi et~al.}{2009}]{Shi2009}
Shi, J., Levinson, D.~F., Duan, J., Sanders, A.~R., Zheng, Y., Pe'er, I.,
  Dudbridge, F., Holmans, P.~A., Whittemore, A.~S., Mowry, B.~J., Olincy, A.,
  Amin, F., Cloninger, C.~R., Silverman, J.~M., Buccola, N.~G., Byerley, W.~F.,
  Black, D.~W., Crowe, R.~R., Oksenberg, J.~R., Mirel, D.~B., Kendler, K.~S.,
  Freedman, R., and Gejman, P.~V. (2009).
\newblock{Common variants on chromosome 6p22.1 are associated with
  schizophrenia}.
\newblock{\em Nature} {\bf 460,} 753--7.

\bibitem[\protect\citeauthoryear{Sullivan, Lin, Tzeng, van~den Oord, Perkins,
  Stroup, Wagner, Lee, Wright, Zou, Liu, Downing, Lieberman, and
  Close}{Sullivan et~al.}{2008}]{Sullivan2008}
Sullivan, P.~F., Lin, D., Tzeng, J.-Y., van~den Oord, E., Perkins, D., Stroup,
  T.~S., Wagner, M., Lee, S., Wright, F.~A., Zou, F., Liu, W., Downing, A.~M.,
  Lieberman, J., and Close, S.~L. (2008).
\newblock{Genomewide association for schizophrenia in the CATIE study: results
  of stage 1}.
\newblock{\em Molecular Psychiatry} {\bf 13,} 570--584.

\bibitem[\protect\citeauthoryear{Wahba}{Wahba}{1990}]{Wahba1990}
Wahba, G. (1990).
\newblock{\em {Spline models for observational data}}.
\newblock Society for Industrial and Applied Mathematics.

\bibitem[\protect\citeauthoryear{Wu, Lee, Cai, Li, Boehnke, and Lin}{Wu
  et~al.}{2011}]{Wu2011}
Wu, M.~C., Lee, S., Cai, T., Li, Y., Boehnke, M., and Lin, X. (2011).
\newblock{Rare-variant association testing for sequencing data with the
  sequence kernel association test.}
\newblock{\em American journal of human genetics} {\bf 89,} 82--93.

\bibitem[\protect\citeauthoryear{Yolken, Torrey, Lieberman, Yang, and
  Dickerson}{Yolken et~al.}{2011}]{Yolken2011}
Yolken, R.~H., Torrey, E.~F., Lieberman, J.~A., Yang, S., and Dickerson, F.~B.
  (2011).
\newblock{Serological evidence of exposure to Herpes Simplex Virus type 1 is
  associated with cognitive deficits in the CATIE schizophrenia sample}.
\newblock{\em Schizophrenia Research} {\bf 128,} 61--65.

\bibitem[\protect\citeauthoryear{Zhan and Wu}{Zhan and Wu}{2017}]{Zhan2017}
Zhan, X. and Wu, M.~C. (2017).
\newblock{Reader Reaction: A note on testing and estimation in marker-set
  association study using semiparametric quantile regression kernel machine}.
\newblock{\em Biometrics} .

\end{thebibliography}

\newcommand{\newblock}{\hskip .11em plus .33em minus .07em}









\label{lastpage}

\end{document}